# A family of well behaved charge analogues of Durgapal's perfect fluid exact solution in general relativity

Mohammad Hassan Murad · Saba Fatema

**Abstract** This paper presents a new family of interior solutions of Einstein–Maxwell field equations in general relativity for a static spherically symmetric distribution of a charged perfect fluid with a particular form of charge distribution. This solution gives us wide range of parameter, *K,* for which the solution is well behaved hence, suitable for modeling of superdense star. For this solution the gravitational mass of a star is maximized with all degree of suitability by assuming the surface density equal to normal nuclear density, $\rho_{nm} = 2.5 \times 10^{17}$ kg m$^{-3}$. By this model we obtain the mass of the Crab pulsar, $M_{Crab}$, $1.3679 M_\odot$ and radius 13.21 km, constraining the moment of inertia $> 1.61 \times 10^{38}$ kg m$^2$ for the conservative estimate of Crab nebula mass $2M_\odot$. And $M_{Crab} = 1.9645 M_\odot$ with radius 14.38 km constraining the moment of inertia $> 3.04 \times 10^{38}$ kg m$^2$ for the newest estimate of Crab nebula mass, $4.6\ M_\odot$. These results are quite well in agreement with the possible values of mass and radius of Crab pulsar. Besides this, our model yields moments of inertia for PSR J0737-3039A and PSR J0737-3039B, $I_A = 1.4285 \times 10^{38}$ kg m$^2$ and $I_B = 1.3647 \times 10^{38}$ kg m$^2$ respectively. It has been observed that under well behaved conditions this class of solutions gives us the overall maximum gravitational mass of super dense object, $M_{G(max)}$, $4.7487 M_\odot$ with radius $R_{M_{max}} = 15.24$ km, surface redshift 0.9878, charge $7.91 \times 10^{20} C$, and central density $4.31 \rho_{nm}$.

## 1. Introduction

Exact solutions of Einstein–Maxwell field equations are of vital importance in relativistic astrophysics. Relativists and astrophysicists have been trying to construct charged perfect fluid models of superdense objects since the formulation of Einstein's gravitational field equations. Of course neutron stars or any stars are not composed of perfect fluid. But such solutions may be used to make a suitable model of superdense object with charge matter for the numerical study of the stellar structure. The study of charged relativistic fluid spheres attracts the interest of researchers of physics and astrophysics due to some of the following reasons:

➢ A spherical body can remain in equilibrium under its own gravitation and electric repulsion, no internal pressure is necessary, if the matter present in the sphere carries certain modest electric charge density (Bonnor 1965). The problem of the stability of a homogeneous distribution of matter containing a net surface charge was considered by Stettner (Cited in Whitman and Burch 1981). He showed that a fluid sphere of uniform density with a modest surface charge is more stable than the same system without charge. His solution is also stable towards an increase in the net surface charge. The electric charge weakens gravity to the extent of turning it into a repulsive field, as happens in the vicinity of a Reissener–Nordström singularity. Thus the gravitational collapse of a charged fluid sphere to a point singularity may be avoided (de Felice et al. 1995).

M. H. Murad (✉), S. Fatema
Department of Natural Sciences, Daffodil International University
102, Sukrabad, Mirpur Road, Dhaka-1207, Bangladesh.
E-mail: murad@daffodilvarsity.edu.bd
saba@daffodilvarsity.edu.bd

- Charged-dust (CD) models and electromagnetic mass models are expected to provide some clue about structure of an electron (Ivanov 2002; Rahaman et al. 2011; Bijalwan 2011).
- Several fluid spheres which do not satisfy some or all the relevant physical conditions i.e. reality conditions, become relevant when they are charged (See Table 4 in Pant and Rajasekhara 2011).

In this paper, we have obtained a family of new two–parametric well behaved exact solutions of Einstein–Maxwell field equations by considering the metric component, $g_{00}$, of the form $B(1 + Cr^2)^4$ (Durgapal 1982) [or Durg IV metric according to the classification made in Delgaty and Lake 1998] with charge distribution $\frac{Cq^2}{x^2} = \frac{K}{2}x(1+x)^n(1+5x)^{\frac{3}{5}}$. Several nonsingular charged analogues of Durg IV solution was obtained earlier by Pant (2011) with $n$ = 0, Maurya and Gupta (2011b), Pant and Rajasekhara (2011) ($n$ = 0). This paper is a continuation and somewhat generalization of Pant's earlier works. Unfortunately, some erroneous equations (pressure and density gradients) are found in Pant (2011). Those equations might have affected their estimation of maximum gravitational mass of the superdense object. In this work, we shall reconsider Pant's work with normal nuclear matter density $\rho_{nm} = 2.5 \times 10^{17} \text{kg m}^{-3}$ (Malone et al. 1975, Shapiro and Teukolsky 2004, kalogera and Baym 1996, Bejger et al. 2005, Page and Reddy 2006).

Neutron stars are some of the densest manifestations of massive objects in the observable universe (Lattimer and Prakash 2004). Constraining the equation of state (EOS) of dense matter in the interior of neutron stars and the determination of its maximum mass have been remaining the greatest challenge in astrophysics. Because one way of establishing the existence of black hole is to show that the mass of compact dense objects exceeds the upper limit of the neutron stars. So we calculate the maximum gravitational mass of stable non-rotating neutron star, in an EOS—independent way, using normal nuclear matter density equal to the surface density of neutron star, following the same algorithm followed earlier by Pant (2011), Murad (2012), and others, with some physical requirements.

In the last two sections we test the compatibility and the well behaved nature of our solution by constructing fluid spheres of various masses with the different values of parameters involved in the solution. We apply this solution to calculate various physical parameters for the rotating compact objects like Crab Pulsar (PSR B0531-21, spin period $P$ = 33 ms), the first double pulsar system PSR J0737-3039 (Burgay et al. 2003; Lyne et al. 2004), composed of two active radio pulsars PSR J0737-3039A ($P$ = 22.69 ms) and PSR J0737-3039B ($P$ = 2.77 s) having precise gravitational mass (1.3381 ± 0.0007) $M_\odot$ and (1.2489 ± 0.0007) $M_\odot$ respectively (Kramer et al. 2006a; 2006b). We find that our solution yields values that are quite well in agreement with the observations made in several recent researches.

## 2. Physical conditions for a regular and well behaved charged fluid sphere to construct a superdense star model

A static spherically symmetric metric in curvature coordinates can be written as (Weinberg 1972)

$$ds^2 = e^{\nu(r)}dt^2 - e^{\lambda(r)}dr^2 - r^2(d\theta^2 + \sin^2\theta\, d\phi^2) \tag{2.1}$$

The functions $\nu$ and $\lambda$ satisfy the Einstein-Maxwell field equation

$$G^i{}_j = R^i{}_j - \frac{1}{2}R\delta^i{}_j = \kappa(T^i{}_j + E^i{}_j) \tag{2.2}$$

where we have chosen the units so that, $c = G = 1$, and $\kappa (= 8\pi)$ is *Einstein's constant*. $T^i{}_j$ and $E^i{}_j$ are the energy-momentum tensor of perfect fluid and electromagnetic field defined by,

$$T^i{}_j = (\rho + p)v^i v_j - p\delta^i{}_j$$
$$E^i{}_j = \frac{1}{4\pi}\left(-F^{im}F_{jm} + \frac{1}{4}\delta^i{}_j F^{mn}F_{mn}\right)$$

where $\rho$, $p$, $v^i$, $F_{ij}$ denote energy density, fluid pressure, velocity vector and anti-symmetric electromagnetic field strength tensor respectively. On account of the high nonlinearity of Einstein–Maxwell field equations not many realistic well behaved analytic solutions are known for the description of relativistic fluid spheres (Stephani et al. 2003; Delgaty and Lake 1998; Lake 2003). For a well behaved model of a relativistic star with charged perfect fluid matter, following physical conditions should be satisfied (Sabbadini and Hartle 1973; Glass and Goldman 1978; Hartle 1978; Buchdahl 1979; Durgapal et al. 1984; Delgaty and Lake 1998; Lake 2003):

(i) The solution should be free from physical and geometric singularities i.e. $e^\nu > 0$ and $e^\lambda > 0$ in the range $0 \leq r \leq R$

(ii) The pressure $p$, at zero temperature is a function of one parameter, $\rho$, only, i.e. $p = p(\rho)$.

(iii) The pressure and density are positive, $p, \rho \geq 0$, where the last inequality is the statement that gravity is attractive.

(iv) Pressure $p$ should be zero at boundary $r = R$.

(v) In order to have an equilibrium configuration the matter must be stable against the collapse of local regions. This requires, *Le Chatelier's principle* also known as *local* or *microscopic stability* condition, that $p$ must be a monotonically non-decreasing function of $\rho$ (Rhoades and Ruffini 1974; Hegyi et al. 1975), mathematically reads,

$$\frac{dp}{d\rho} \geq 0$$

(vi) The quantity $\sqrt{\frac{dp}{d\rho}}$ is the hydrodynamic phase velocity of sound waves in the neutron star matter. In the absence of dispersion and absorption it would be the velocity of signals in the medium. Then the condition $\sqrt{\frac{dp}{d\rho}} \leq 1$ would then be the condition that the speed of these signals not exceeds that of light (*causality condition*).

(vii) $\rho \geq p$, the dominant energy condition.

(viii) The trace of the energy momentum tensor must be nonnegative, i.e., $\rho - 3p \geq 0$, $0 \leq r < R$.

(ix) $\left(\frac{dp}{dr}\right)_{r=0}, \left(\frac{d\rho}{dr}\right)_{r=0} = 0$ and $\left(\frac{d^2p}{dr^2}\right)_{r=0}, \left(\frac{d^2\rho}{dr^2}\right)_{r=0} < 0$ so that pressure and density gradients $\frac{dp}{dr}, \frac{d\rho}{dr} < 0$ for $0 < r \leq R$. In other words, the conditions imply that pressure and density should maximum at the centre and monotonically decreasing towards the pressure free interface (i.e. boundary of the fluid sphere).

(x) The velocity of sound should be decreasing towards the surface i.e. $\left(\frac{d}{dr}\left(\frac{dp}{d\rho}\right)\right)_{r=0} < 0$ for $0 \leq r \leq R$ or the velocity of sound is increasing with the increase of density.

(xi) The ratio of pressure to the density $\frac{p}{\rho}$ should be monotonically decreasing with the increase of $r$ i.e. $\left(\frac{d}{dr}\left(\frac{p}{\rho}\right)\right)_{r=0} < 0$ and $\left(\frac{d^2}{dr^2}\left(\frac{p}{\rho}\right)\right)_{r=0} < 0$ and $\frac{d}{dr}\left(\frac{p}{\rho}\right)$ is negative valued function for $0 \leq r \leq R$.

(xii) Gravitational redshift, $z$, should be monotonically decreasing toward the boundary of the sphere. The central red shift, $z_c$, and surface red shift, $z_R$, should be positive and finite.

(xiii) Electric field intensity $E$, such that $E(0) = 0$, is taken to be monotonically increasing i.e. $\frac{dE}{dr} > 0$ for $0 < r < R$.

(xiv) The relativistic adiabatic index is given by $\gamma = \frac{(p+\rho)}{p}\frac{dp}{d\rho}$. The necessary condition for this exact solution to serve as a model of a relativistic star is that $\gamma > \frac{4}{3}$ (Ipser 1970; Adams et al. 1975; Knutsen 1988a; 1988b; 1989; 1991).

Buchdahl (1959) has obtained an absolute constraint of the maximally allowable mass–radius (M/R) ratio for isotropic fluid spheres of the form $\frac{2M}{R} \leq \frac{8}{9}$ (in natural units, $c = G = 1$) which states that, for a given radius a static isotropic fluid sphere cannot have an arbitrary mass. Böhmer and Harko (2007) proved that for a compact object with charge, $Q$ ($< M$), there is a lower bound for the mass–radius ratio,

$$\frac{3}{2}\frac{Q^2}{R^2}\frac{\left(1 + \frac{Q^2}{18R^2}\right)}{\left(1 + \frac{Q^2}{12R^2}\right)} \leq \frac{2M}{R}$$

Andréasson (2009) generalized the Buchdahl inequality for the charged case,

$$\sqrt{M} \leq \frac{\sqrt{R}}{3} + \sqrt{\frac{R}{9} + \frac{Q^2}{3R}}$$

Combining these results to constrain the mass-to-radius ratio, reads,

$$\frac{3}{2}\frac{Q^2}{R^2}\frac{\left(1+\frac{Q^2}{18R^2}\right)}{\left(1+\frac{Q^2}{12R^2}\right)} \leq \frac{2M}{R} \leq 2\left(\frac{R+\sqrt{(R^2+3Q^2)}}{3R}\right)^2$$

In the forthcoming sections we shall be using the following theorem (Pant et al. 2011) for showing the monotonically decreasing or increasing nature of various physical parameters for well behaved nature of our solution.

**Theorem 2.1** If $f(r) = g(x)$; $\left(\frac{dg}{dx}\right)_{x=0}$ and $\left(\frac{d^2g}{dx^2}\right)_{x=0}$ are non zero finite, where $x = Cr^2, C > 0$, then,

Maxima of $f(r)$ exists at $r = 0$ if $\left(\frac{dg}{dx}\right)_{x=0}$ is finitely negative

Minima of $f(r)$ exists at $r = 0$ if $\left(\frac{dg}{dx}\right)_{x=0}$ is finitely positive

## 3. Einstein–Maxwell equation for charged perfect fluid distribution

In view of the metric (2.1), the field equation (2.2) gives (Nduka 1976; Singh and Yadav 1978; Dionysiou 1982),

$$\frac{\nu'}{r}e^{-\lambda} - \frac{(1-e^{-\lambda})}{r^2} = \kappa p - \frac{q^2}{r^4} \tag{3.1}$$

$$\left(\frac{\nu''}{2} - \frac{\nu'\lambda'}{4} + \frac{\nu'^2}{4} + \frac{\nu'-\lambda'}{2r}\right)e^{-\lambda} = \kappa p + \frac{q^2}{r^4} \tag{3.2}$$

$$\frac{\lambda'}{r}e^{-\lambda} + \frac{(1-e^{-\lambda})}{r^2} = \kappa c^2 \rho + \frac{q^2}{r^4} \tag{3.3}$$

where, prime ($\lambda'$) denotes the differentiation with respect to $r$ and $q(r)$ represents the total charge contained within the sphere of radius $r$.
Now let us assume

$$\left. \begin{array}{l} e^\nu = B(1+Cr^2)^4 \\ e^{-\lambda} = Z, \ x = Cr^2 \end{array} \right\} \tag{3.4}$$

Putting these transformations into (3.1) and (3.3), the equations become,

$$\frac{\kappa}{C}p = \frac{(1+9x)}{x(1+x)}Z - \frac{1}{x} + \frac{Cq^2}{x^2} \tag{3.5}$$

$$\frac{\kappa}{C}\rho = -2\frac{dZ}{dx} - \frac{Z}{x} - \frac{1}{x}\left(\frac{Cq^2}{x}-1\right) \tag{3.6}$$

and (3.2) becomes,

$$\frac{dZ}{dx} - \frac{(7x^2-2x-1)}{x(1+x)(1+5x)}Z = \frac{(1+x)}{x(1+5x)}\left(\frac{2Cq^2}{x}-1\right) \tag{3.7}$$

The solution is

$$Z\frac{(1+x)^2(1+5x)^{\frac{2}{5}}}{x} = \int\left\{\frac{(1+x)^3}{x^2(1+5x)^{\frac{3}{5}}}\left(\frac{2Cq^2}{x}-1\right)\right\}dx + A \tag{3.8}$$

where $A$ is an arbitrary constant of integration.

To integrate (3.8) we assume,

$$\frac{E^2}{C} = \frac{Cq^2}{x^2} = \frac{K}{2}x(1+x)^n(1+5x)^{\frac{3}{5}} \tag{3.9}$$

where $K \geq 0$. The electric intensity is so assumed that the model is physically significant and well behaved. In view of (3.9), (3.8) yields the following solution,

$$Z = \frac{K}{n+4} \frac{x(1+x)^{n+2}}{(1+5x)^{\frac{2}{5}}} + \frac{(7-10x-x^2)}{7(1+x)^2} + A \frac{x}{(1+x)^2(1+5x)^{\frac{2}{5}}} \qquad (3.10)$$

Using (3.9) and (3.10) into (3.5) and (3.6), the pressure and energy density become,

$$\frac{\kappa}{C}p = \frac{K(1+x)^n}{2(n+4)}\left[\frac{2+(n+24)x+(5n+38)x^2}{(1+5x)^{\frac{2}{5}}}\right] + \frac{16}{7}\frac{(2-7x-x^2)}{(1+x)^3} + A\frac{(1+9x)}{(1+x)^3(1+5x)^{\frac{2}{5}}} \qquad (3.11)$$

and

$$\frac{\kappa}{C}\rho = -\frac{K(1+x)^n}{2(n+4)(1+5x)^{\frac{7}{5}}}\left[\begin{matrix}6+(5n+46)x+(34n+138)x^2\\+(45n+162)x^3\end{matrix}\right] + \frac{8}{7}\frac{(9+2x+x^2)}{(1+x)^3}$$

$$-A\frac{(3+10x-9x^2)}{(1+x)^3(1+5x)^{\frac{7}{5}}} \qquad (3.12)$$

## 4. Properties of new class of solution

The central values of pressure and density are given by,

$$\frac{\kappa}{C}p_c = \frac{K}{(n+4)} + \frac{32}{7} + A$$

$$\frac{\kappa}{C}\rho_c = -\frac{3K}{(n+4)} + \frac{72}{7} - 3A$$

For $p_c$ and $\rho_c$ must be positive and $\frac{p_c}{\rho_c} \leq 1$, we have,

$$-\frac{K}{(n+4)} - \frac{32}{7} \leq A \leq -\frac{K}{(n+4)} + \frac{10}{7} \qquad (4.1)$$

Differentiating (3.11) and (3.12) with respect to $x$, we obtain the pressure and density gradients,

$$\frac{\kappa}{C}\frac{dp}{dx} = \frac{K}{2(n+4)}\frac{(1+x)^{n-1}}{(1+5x)^{\frac{7}{5}}}P_n(x) + \frac{16}{7}\frac{(-13+12x+x^2)}{(1+x)^4} + 4A\frac{(1-2x-27x^2)}{(1+x)^4(1+5x)^{\frac{7}{5}}} \qquad (4.2)$$

$$\frac{\kappa}{C}\frac{d\rho}{dx} = -\frac{K}{2(n+4)}\frac{(1+x)^{n-1}}{(1+5x)^{\frac{12}{5}}}Q_n(x) - \frac{8}{7}\frac{(25+2x+x^2)}{(1+x)^4} + 4A\frac{(5+31x+47x^2-27x^3)}{(1+x)^4(1+5x)^{\frac{12}{5}}} \qquad (4.3)$$

where,

$$P_n(x) = \begin{matrix}(3n+20)+(n^2+48n+168)x+(10n^2+211n+452)x^2\\+(25n^2+230n+304)x^3\end{matrix}$$

$$Q_n(x) = \begin{matrix}(11n+4)+(5n^2+139n+188)x+(59n^2+663n+1084)x^2+\\(215n^2+1449n+2196)x^3+(225n^2+1170n+1296)x^4\end{matrix}$$

And for the values of $K \geq 0$ and $A$ satisfied by (4.1) the following must be satisfied

$$0 \leq \left(\frac{dp}{d\rho}\right)_{x=0} \leq 1$$

$$\frac{d}{dx}\left(\frac{p}{\rho}\right)_{x=0} < 0$$

and

$$\frac{d}{dx}\left(\frac{E^2}{C}\right)_{x=0} = \frac{K}{2} \geq 0$$

$$\frac{d^2}{dx^2}\left(\frac{E^2}{C}\right)_{x=0} = (n+3)K \neq 0;\ n = 0, 1, 2 \ldots$$

By Theorem 2.1, the above two inequalities show that electric field intensity $E$ is minimum at the centre and monotonically increasing towards the boundary of the sphere.

## 5. Physical Boundary Conditions

Besides the above, the charged fluid spheres so obtained are to be matched over the boundary with Reissner-Nordström metric (Dionysiou 1982):

$$ds^2 = \left(1 - \frac{2M}{r} + \frac{q^2}{r^2}\right)dt^2 - \left(1 - \frac{2M}{r} + \frac{q^2}{r^2}\right)^{-1}dr^2 - r^2(d\theta^2 + \sin^2\theta\, d\phi^2) \tag{5.1}$$

which requires the continuity of $e^\nu$, $e^{-\lambda}$ and $q$ across the boundary $r = R$

$$e^{\nu(R)} = \left(1 - \frac{2M}{R} + \frac{Q^2}{R^2}\right) \tag{5.2}$$

$$e^{-\lambda(R)} = \left(1 - \frac{2M}{R} + \frac{Q^2}{R^2}\right) \tag{5.3}$$

$$q(R) = Q$$

$$p(R) = 0$$

where $M$, $R$ and $Q$ represent the total mass, radius and the total charge inside the fluid sphere respectively.

Now using $r = R, x = CR^2 = X$ and $p(R) = 0$ into (3.11) we can compute the arbitrary constant $A$

$$A = -\frac{K(1+X)^{n+3}}{2(n+4)}\left[\frac{2 + (n+24)X + (5n+38)X^2}{(1+9X)}\right] - \frac{16}{7}\frac{(2 - 7X - X^2)(1+5X)^{\frac{2}{5}}}{(1+9X)} \tag{5.4}$$

Using (5.2) and (3.4) we can construct the constant, $B$, as

$$B = \frac{K}{(n+4)}\frac{X(1+X)^{n-2}}{(1+5X)^{\frac{2}{5}}} + \frac{(7 - 10X - X^2)}{7(1+X)^6} + A\frac{X}{(1+X)^6(1+5X)^{\frac{2}{5}}} \tag{5.5}$$

The expression for the gravitational redshift is given by

$$z = \sqrt{e^{-\nu}} - 1 = \frac{1}{\sqrt{B}(1+x)^2} - 1$$

The central red shift is given by

$$z_c = \sqrt{e^{-\nu(0)}} - 1 = \frac{1}{\sqrt{B}} - 1$$

The condition $z_c > 0$ implies,

$$0 < \sqrt{B} < 1$$

and

$$\left(\frac{dz}{dx}\right)_{x=0} = -\frac{2}{\sqrt{B}} < 0$$

$$\left(\frac{d^2z}{dx^2}\right)_{x=0} = \frac{6}{\sqrt{B}} \neq 0$$

By Theorem 2.1, the above two inequalities show that gravitational redshift $z$ is maximum at the centre and monotonically decreasing towards the boundary of the fluid sphere.

Denoting the boundary surface density $\rho(R) = \rho_s$, (3.12) gives,

$$\kappa R^2 \rho_s = -\frac{KX(1+X)^n}{2(n+4)(1+5X)^{\frac{7}{5}}}\begin{bmatrix} 6 + (5n+46)X + (34n+138)X^2 \\ +(45n+162)X^3 \end{bmatrix} + \frac{8X}{7}\frac{(9+2X+X^2)}{(1+X)^3}$$

$$-AX\frac{(3+10X-9X^2)}{(1+X)^3(1+5X)^{\frac{7}{5}}} = L \tag{5.7}$$

Now the radius of the charged fluid sphere becomes,

$$R = \sqrt{\frac{L}{\kappa \rho_s}} \tag{5.8}$$

Using (5.2), (3.4), and (5.8) we can construct the mass-to-radius ratio, which reads,

$$\frac{2M}{R} = \frac{8X(3+X)}{7(1+X)^2} + \frac{KX(1+X)^n}{2(n+4)}\left(\frac{-2+nX+(5n+18)X^2}{(1+5X)^{\frac{2}{5}}}\right) - A\frac{X}{(1+X)^2(1+5X)^{\frac{2}{5}}} \tag{5.5}$$

## 6. Calculations and Tables of numerical values

To construct well behaved model of superdense astrophysical object we shall be using the following numerical values

Nuclear matter density, $\rho_{nm} = 1.857 \times 10^{-10}$ m$^{-2}$ = $2.5 \times 10^{17}$ kg m$^{-3}$,
Velocity of light, $c = 1 = 2.997 \times 10^8$ ms$^{-1}$,
Newton's gravitational constant, $G = 1 = 6.674 \times 10^{-11}$ Nm$^2$kg$^{-2}$,
Mass of the Sun, $M_\odot = 1.486$ km $= 2 \times 10^{30}$ kg

**Table 1** The variation of various physical parameters e. g. pressure, surface density, pressure-energy density ratio, causality condition, surface redshift, charge-to-mass ratio.

| $n$ | $K$ | $X$ | $\frac{\kappa}{C}p_c$ | $\frac{\kappa}{C}c^2\rho_c$ | $\frac{1}{c^2}\left(\frac{p}{\rho}\right)_c$ | $\sqrt{\frac{1}{c^2}\left(\frac{dp}{d\rho}\right)_c}$ | $z_R$ | $\frac{M_G}{M_\odot}$ | $R$ (km) | $\frac{Q \text{ (km)}}{M_G \text{(km)}}$ | $\rho_{c,17}$ |
|---|---|---|---|---|---|---|---|---|---|---|---|
| 0 | 2.03 | 0.052 | 1.6152 | 19.1543 | 0.0843 | 0.6240 | 0.1831 | 1.2496 | 12.86 | 0.38 | 3.22 |
| 0 | 2 | 0.057 | 1.7217 | 18.8348 | 0.0914 | 0.6282 | 0.1986 | 1.3679 | 13.21 | 0.39 | 3.29 |
| 0 | 1.23 | 0.39 | 3.9988 | 12.0035 | 0.3331 | 0.8028 | 0.9879 | 4.7487[a] | 15.24 | 0.91 | 10.79 |
| 0 | 3.8 | 0.055 | 1.5238 | 19.4285 | 0.0784 | 0.5726 | 0.1943 | 1.3380 | 13.01 | 0.53 | 3.38 |
| 0 | 3.76 | 0.083 | 1.9350 | 18.1949 | 0.1063 | 0.5826 | 0.2803 | 1.9645 | 14.38 | 0.62 | 3.90 |
| 0 | 13.8 | 0.054 | 0.6505 | 22.0484 | 0.0295 | 0.3674 | 0.2014 | 1.4004 | 12.59 | 0.92 | 4.01 |
| 1 | 0.97 | 0.363 | 3.9971 | 12.0084 | 0.3328 | 0.8204 | 0.9329 | 4.6456[b] | 15.43 | 0.90 | 9.79 |
| 1 | 13.1 | 0.054 | 0.6700 | 21.9898 | 0.0304 | 0.3790 | 0.2014 | 1.4004 | 12.59 | 0.92 | 4.00 |
| 1 | 1.87 | 0.057 | 1.7270 | 18.8188 | 0.0917 | 0.6323 | 0.1986 | 1.3677 | 13.21 | 0.39 | 3.29 |
| 1 | 3.6 | 0.055 | 1.5297 | 19.4107 | 0.0788 | 0.5779 | 0.1943 | 1.3380 | 13.01 | 0.53 | 3.37 |
| 2 | 3.41 | 0.055 | 1.5355 | 19.3934 | 0.0791 | 0.5831 | 0.1943 | 1.3380 | 13.01 | 0.53 | 3.37 |
| 2 | 12.4 | 0.054 | 0.6921 | 21.9234 | 0.0315 | 0.3908 | 0.2013 | 1.4002 | 12.59 | 0.92 | 3.99 |
| 2 | 0.77 | 0.343 | 3.9960 | 12.0119 | 0.3326 | 0.8345 | 0.8907 | 4.5573[c] | 15.58 | 0.88 | 9.08 |
| 5 | 0.41 | 0.3 | 3.9642 | 12.1072 | 0.3274 | 0.8565 | 0.7970 | 4.3275[d] | 15.89 | 0.85 | 7.69 |
| 10 | 0.14 | 0.271 | 3.9732 | 12.0803 | 0.3289 | 0.8787 | 0.7286 | 4.1267[e] | 16.14 | 0.81 | 6.72 |
| 15 | 0.05 | 0.254 | 3.9700 | 12.0898 | 0.3283 | 0.8854 | 0.6864 | 3.9863[f] | 16.29 | 0.77 | 6.19 |
| 20 | 0.13 | 0.165 | 3.1442 | 14.5672 | 0.2158 | 0.7805 | 0.5014 | 3.2454[g] | 15.80 | 0.76 | 5.15 |
| 25 | 0.13 | 0.138 | 2.8567 | 15.4297 | 0.1851 | 0.7564 | 0.4347 | 2.9100[h] | 15.52 | 0.74 | 4.73 |

| | | | | | | | | | | |
|---|---|---|---|---|---|---|---|---|---|---|
| 30 | 0.12 | 0.121 | 2.6565 | 16.0302 | 0.1657 | 0.7421 | 0.3903 | 2.6634[i] | 15.26 | 0.73 | 4.45 |
| 40 | 0.099 | 0.099 | 2.3644 | 16.9067 | 0.1398 | 0.7239 | 0.3297 | 2.2960[j] | 14.78 | 0.71 | 4.09 |
| 50 | 0.08 | 0.085 | 2.1587 | 17.5236 | 0.1232 | 0.7127 | 0.2891 | 2.0278[h] | 14.37 | 0.69 | 3.86 |

Notations, $\rho_{c,17} = \dfrac{\rho_c}{10^{17}\text{kg m}^{-3}}$

[a] The maximum gravitational mass, $M_{G(\max)}$, of the superdense object with $n = 0$.

[b] $M_{G(\max)}$ with $n = 1$.

[c] $M_{G(\max)}$ with $n = 2$.

[d] $M_{G(\max)}$ with $n = 5$.

[e] $M_{G(\max)}$ with $n = 10$.

[f] $M_{G(\max)}$ with $n = 15$.

[g] $M_{G(\max)}$ with $n = 20$.

[h] $M_{G(\max)}$ with $n = 25$.

[i] $M_{G(\max)}$ with $n = 30$.

[j] $M_{G(\max)}$ with $n = 40$.

[h] $M_{G(\max)}$ with $n = 50$.

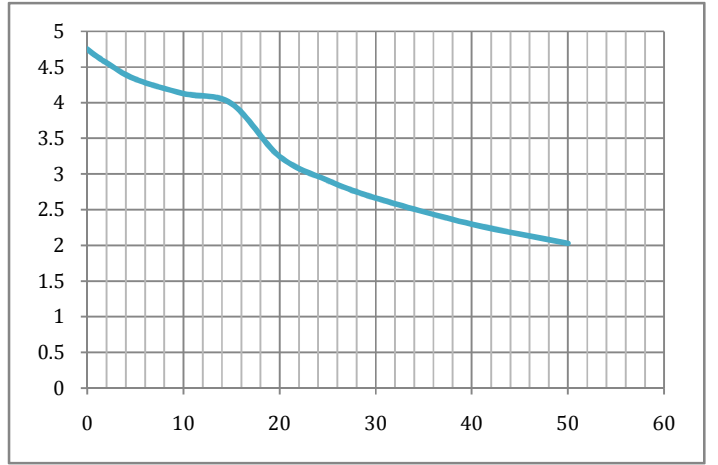

**Figure 1**: The variation of maximum gravitational mass (in the unit of km) with the variation of charge. [$n$ and $M_G$ are plotted along the horizontal and vertical axes respectively.]

**Table 2** The variation of various physical parameters e. g. pressure, surface density, pressure-energy density ratio, causality condition, gravitational redshift, pressure gradient, density gradient, charge, and relativistic adiabatic index in the fluid sphere with $n = 0$, $K = 1.23$, $X = 0.39$

| $\dfrac{r}{R}$ | $R^2\kappa p$ | $R^2\kappa c^2\rho$ | $\dfrac{1}{c^2}\dfrac{p}{\rho}$ | $\dfrac{\kappa}{C}\dfrac{dp}{dx}$ | $\dfrac{\kappa c^2}{C}\dfrac{d\rho}{dx}$ | $\sqrt{\dfrac{1}{c^2}\dfrac{dp}{d\rho}}$ | $z$ | $\gamma$ | $Q$ (km) |
|---|---|---|---|---|---|---|---|---|---|
| 0 | 3.998806 | 12.00358 | 0.333134 | -30.1598 | -46.7889 | 0.802865426 | 2.840995 | 3.2129 | 0 |
| 0.1 | 3.88264 | 11.82308 | 0.328395 | -29.4158 | -45.7813 | 0.801578309 | 2.81121 | 3.242476 | 0.046888 |
| 0.2 | 3.551036 | 11.30411 | 0.314137 | -27.296 | -42.9847 | 0.796878469 | 2.723903 | 3.333604 | 0.190719 |
| 0.3 | 3.050563 | 10.50635 | 0.290354 | -24.1007 | -38.9556 | 0.786556534 | 2.584917 | 3.495511 | 0.440411 |
| 0.4 | 2.446895 | 9.50832 | 0.257343 | -20.2313 | -34.3301 | 0.767669209 | 2.403044 | 3.750733 | 0.809188 |
| 0.5 | 1.811833 | 8.389213 | 0.215972 | -16.0948 | -29.6336 | 0.736970095 | 2.188854 | 4.149315 | 1.312884 |
| 0.6 | 1.211884 | 7.216959 | 0.167922 | -12.0344 | -25.2182 | 0.690804201 | 1.953448 | 4.804651 | 1.968519 |
| 0.7 | 0.7007 | 6.043033 | 0.115952 | -8.29492 | -21.2811 | 0.624323062 | 1.707368 | 6.008661 | 2.793332 |
| 0.8 | 0.315877 | 4.901794 | 0.064441 | -5.01821 | -17.9072 | 0.529371662 | 1.459811 | 8.744182 | 3.804241 |
| 0.9 | 0.07934 | 3.812283 | 0.020812 | -2.25806 | -15.1077 | 0.386605708 | 1.218186 | 18.96302 | 5.017586 |
| 1 | 0 | 2.781135 | 0 | -0.0039 | -12.849 | 0.017416073 | 0.98799 | ∞ | 6.449051 |

**Table 3** The variation of various physical parameters e. g. pressure, surface density, pressure-energy density ratio, causality condition, gravitational redshift, pressure gradient, density gradient, charge, and relativistic adiabatic index in the fluid sphere with $n = 1$, $K = 0.97$, $X = 0.363$

| $\frac{r}{R}$ | $R^2\kappa p$ | $R^2\kappa c^2\rho$ | $\frac{1}{c^2}\frac{p}{\rho}$ | $\frac{\kappa}{C}\frac{dp}{dx}$ | $\frac{\kappa c^2}{C}\frac{d\rho}{dx}$ | $\sqrt{\frac{1}{c^2}\frac{dp}{d\rho}}$ | $z$ | $\gamma$ | $Q$ (km) |
|---|---|---|---|---|---|---|---|---|---|
| 0 | 3.997173 | 12.00848 | 0.332863 | -30.5563 | -45.3915 | 0.820470544 | 2.590911 | 3.285364 | 0 |
| 0.1 | 3.887507 | 11.84532 | 0.328189 | -29.8687 | -44.5086 | 0.819192582 | 2.564982 | 3.315291 | 0.039308 |
| 0.2 | 3.573085 | 11.37428 | 0.314137 | -27.8992 | -42.0455 | 0.814583403 | 2.488859 | 3.407664 | 0.160574 |
| 0.3 | 3.09444 | 10.64453 | 0.290707 | -24.8998 | -38.4664 | 0.804557438 | 2.367298 | 3.572146 | 0.373499 |
| 0.4 | 2.509746 | 9.721675 | 0.25816 | -21.2125 | -34.3167 | 0.786218499 | 2.207507 | 3.83169 | 0.693345 |
| 0.5 | 1.884389 | 8.672809 | 0.217275 | -17.1921 | -30.0659 | 0.756184253 | 2.018243 | 4.236486 | 1.139993 |
| 0.6 | 1.281479 | 7.556138 | 0.169594 | -13.1464 | -26.0464 | 0.710442996 | 1.808829 | 4.899513 | 1.737154 |
| 0.7 | 0.755243 | 6.41575 | 0.117717 | -9.30307 | -22.4607 | 0.643577789 | 1.588272 | 6.110737 | 2.511835 |
| 0.8 | 0.347883 | 5.28016 | 0.065885 | -5.80168 | -19.4125 | 0.546683096 | 1.3646 | 8.844232 | 3.494033 |
| 0.9 | 0.089478 | 4.163036 | 0.021493 | -2.70237 | -16.9375 | 0.399436652 | 1.14445 | 18.9835 | 4.716576 |
| 1 | 0 | 3.065041 | 0 | -0.00388 | -15.0278 | 0.01606556 | 0.932916 | $\infty$ | 6.215043 |

**Table 4** The variation of various physical parameters e. g. pressure, surface density, pressure-energy density ratio, causality, gravitational redshift, pressure gradient, density gradient, charge, and relativistic adiabatic index in the fluid sphere with $n = 10$, $K = 0.14$, $X = 0.271$

| $\frac{r}{R}$ | $R^2\kappa p$ | $R^2\kappa c^2\rho$ | $\frac{1}{c^2}\frac{p}{\rho}$ | $\frac{\kappa}{C}\frac{dp}{dx}$ | $\frac{\kappa c^2}{C}\frac{d\rho}{dx}$ | $\sqrt{\frac{1}{c^2}\frac{dp}{d\rho}}$ | $z$ | $\gamma$ | $Q$ (km) |
|---|---|---|---|---|---|---|---|---|---|
| 0 | 3.973212 | 12.08036 | 0.328898 | -31.8972 | -41.3058 | 0.87876075 | 1.792566 | 3.550591 | 0 |
| 0.1 | 3.887436 | 11.96923 | 0.324786 | -31.4076 | -40.7113 | 0.878334184 | 1.777492 | 3.582684 | 0.011778 |
| 0.2 | 3.637918 | 11.6452 | 0.312396 | -29.9859 | -39.0257 | 0.876562677 | 1.732994 | 3.682494 | 0.049634 |
| 0.3 | 3.246875 | 11.13384 | 0.291622 | -27.7614 | -36.5085 | 0.87201344 | 1.661171 | 3.86223 | 0.121617 |
| 0.4 | 2.747706 | 10.47065 | 0.26242 | -24.914 | -33.504 | 0.862330076 | 1.565282 | 4.148401 | 0.242823 |
| 0.5 | 2.180844 | 9.69319 | 0.224987 | -21.633 | -30.3729 | 0.843948042 | 1.449426 | 4.59504 | 0.438258 |
| 0.6 | 1.59008 | 8.833395 | 0.180008 | -18.0731 | -27.4672 | 0.811164476 | 1.318179 | 5.317439 | 0.747549 |
| 0.7 | 1.020771 | 7.909846 | 0.129051 | -14.3091 | -25.1545 | 0.754219452 | 1.176229 | 6.598583 | 1.232304 |
| 0.8 | 0.521577 | 6.918067 | 0.075394 | -10.2827 | -23.8921 | 0.656034989 | 1.028064 | 9.357514 | 1.987349 |
| 0.9 | 0.152005 | 5.814699 | 0.026142 | -5.72346 | -24.3646 | 0.484674274 | 0.877728 | 19.02508 | 3.157607 |
| 1 | 0 | 4.487989 | 0 | -0.00997 | -27.726 | 0.018961569 | 0.728671 | $\infty$ | 4.963152 |

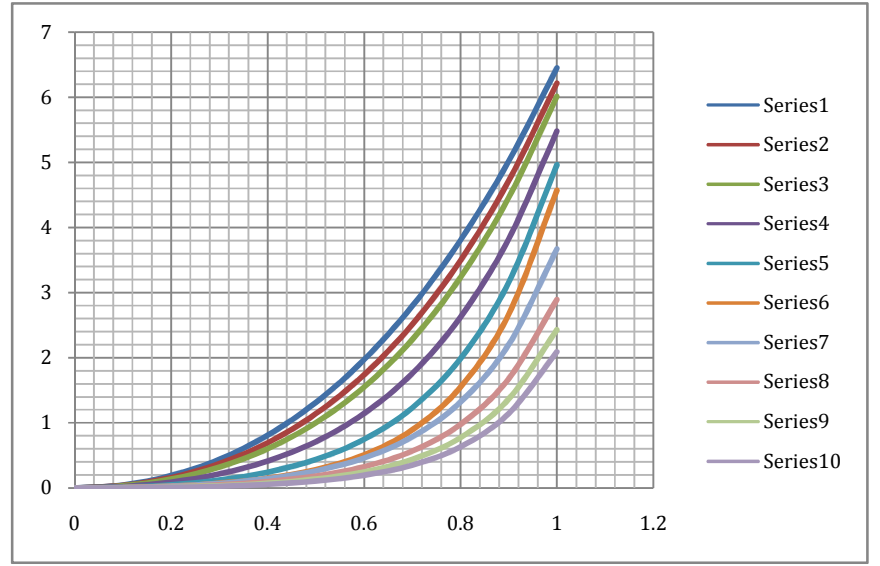

**Figure 2**: The variation of charge (in the unit of km) within the fluid sphere [The fractional radius $\left(\frac{r}{R}\right)$ and charge $Q$ are plotted along the horizontal and vertical axes respectively.]

Series1 ($n, K, X$) = (0, 1.23, 0.39)
Series2 ($n, K, X$) = (1, 1.04, 0.35)
Series3 ($n, K, X$) = (2, 0.77, 0.343)
Series4 ($n, K, X$) = (5, 0.41, 0.3)
Series5 ($n, K, X$) = (10, 0.14, 0.271)
Series6 ($n, K, X$) = (15, 0.05, 0.254)
Series7 ($n, K, X$) = (20, 0.13, 0.165)
Series8 ($n, K, X$) = (30, 0.12, 0.121)
Series9 ($n, K, X$) = (40, 0.099, 0.099)
Series10 ($n, K, X$) = (50, 0.08, 0.085)

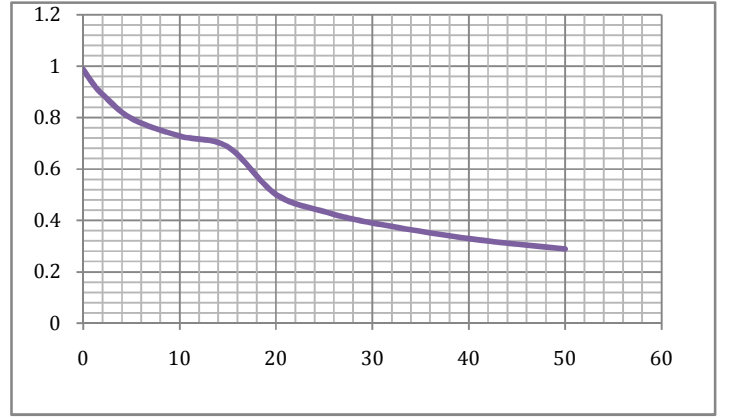

**Figure 3**: The variation of surface redshift with the variation of charge [$n$ and $z_R$ are plotted along the horizontal and vertical axes respectively.]

## 7. An application of the model to the various well known pulsars

The mass and the moment of inertia are the two gross structural parameters of neutron stars which are most accessible to observation. It is the mass which controls the gravitational interaction of the star with other systems such as a binary companion. It is the moment of inertia which controls the energy stored in rotation and thereby the energy available to the pulsar emission mechanism. In this section we calculate the moment of inertia for the Crab Pulsar (PSR B0531-21, spin period $P$ = 33 ms), or PSR J0737-3039 composed of two pulsars PSR J0737-3039A ($P$ = 22.69 ms, $M_G$ = 1.338$M_\odot$) and PSR J0737-3039B ($P$ = 2.77 s, $M_G$ = 1.249$M_\odot$), and one of the most recently discovered (Demorest et al. 2010) massive binary millisecond pulsar PSR J1614-2230 ($P$ = 3.15 ms, $M_G$ = 1.97 ± 0.04$M_\odot$) by the very precise, "empirical formula" which is based on the numerical results obtained for a thirty theoretical equations of state (EOS) of dense nuclear matter (Bejger and Haensel 2002),

$$I \simeq a(x) M R^2 \qquad (7.1)$$

$$a(x) = \begin{cases} a_{NS}(x) = \begin{cases} \dfrac{x}{(0.1 + 2x)} & x \leq 0.1 \\ \dfrac{2}{9}(1 + 5x) & x > 0.1 \end{cases} \\ a_{SS}(x) = \dfrac{2}{5}(1 + x) \end{cases} \qquad (7.2)$$

where $x$ is the dimensionless compactness parameter ($M$ and $R$ are measured in km).

$$x = \frac{\left(\frac{M}{R}\right)}{M_\odot} = \frac{\frac{M}{R}}{1.486}$$

Equation (7.1) and (7.2) are used to calculate the moment of inertia of several fluid spheres for the model considered in the present study.

**Table 5** Moments of inertia of various well behaved charged fluid spheres with known gravitational mass of pulsars

| $n$ | $K$ | $X$ | $\frac{M_G}{M_\odot}$ | $R$ (km) | $I_{NS,38}$ | $\frac{\rho_c}{\rho_{nm}}$ | $Q_{20}$ |
|---|---|---|---|---|---|---|---|
| 0 | 2 | 0.057 | $1.3679^a$ | 13.21 | 1.6102 | 1.31 | 2.27 |
| 1 | 1.87 | | $1.3677^a$ | | 1.6100 | | |
| 0 | 3.76 | 0.083 | $1.9645^b$ | 14.38 | 3.0402 | 1.56 | 3.27 |
| 1 | 3.47 | | $1.9644^b$ | | | | |
| 0 | 1.23 | 0.39 | $4.7487^c$ | 15.24 | 12.5415 | 4.31 | 7.91 |
| 1 | 0.97 | 0.363 | $4.6456^d$ | 15.43 | 12.3244 | 3.91 | 7.74 |
| 0 | 5.66 | 0.072 | $1.3381^e$ | 12.59 | 1.4447 | 1.54 | 2.23 |
| 1 | 5.28 | | | | | 1.53 | |
| 0 | 8.43 | 0.05 | $1.2496^f$ | 12.52 | 1.3053 | 1.42 | 2.08 |
| 1 | 8.02 | | | | | 1.41 | |

Notations $I_{NS,38} = \frac{I_{NS}}{10^{38}\,\text{kg m}^2}$, $Q_{20} = \frac{Q\,(\text{in } C)}{10^{20} C}$

[a] The gravitational mass of the Crab pulsar, with different charge variation, constraining the moment of inertia $I_{\text{Crab},38} > 1.61$ for the conservative estimate of Crab Nebula mass $M_{\text{Neb}} = 2M_\odot$ (Bejger and Haensel 2002; Murad 2012).

[b] The gravitational mass of the Crab pulsar, with different charge variation, constraining the moment of inertia $I_{\text{Crab},38} > 3.04$ for the newest estimate Crab Nebula mass $M_{\text{Neb}} = 4.6 \pm 1.8 M_\odot$ (Bejger and Haensel 2002; Murad 2012).

[c] $M_{G(\max)}$ with $n = 0$.

[d] $M_{G(\max)}$ with $n = 1$.

[e] The gravitational mass of the pulsar PSR J0737-3039A (Kramer et al. 2006a; 2006b)

[f] The gravitational mass of the pulsar PSR J0737-3039B (Kramer et al. 2006a; 2006b)

## 8. Discussions and Conclusion

In view of Table 1, we observe that all the physical parameters $\left(p, \rho, \frac{p}{\rho}, \frac{dp}{d\rho}, z, E\right)$ are positive at the centre and within the limit of realistic EOS. In this article, the well behaved Durgapal IV solution of Einstein's gravitational field equations in general relativity has been charged by means of suitable electric charge distribution, which is zero at the center and monotonically increasing towards the pressure free interface. Our solution satisfies well behaved conditions only for wide range of values of $K$. It also has been observed that the resulting charged fluid spheres can be utilized to construct superdense star models of various compact astrophysical charged objects such as neutron star or black hole. Owing to the various conditions that we obtain here we arrive at the conclusion that under well behaved conditions this class of solutions gives us the gravitational mass of superdense objects. Corresponding to the values ($n$, $K$, $X$) = (0, 1.23, 0.39), we found the overall maximum gravitational mass $M_{G(\max)} = 4.7487 M_\odot$. This mass, however, exceeds the upper limit of maximum mass of equilibrium configuration of a non-rotating neutron star calculated earlier by Rhoades and Ruffini (1974, $3.2 M_\odot$), and Malone et al. (1975, $2 M_\odot$). Recently Rhoades–Ruffini upper limit has been modified by Kalogera and Baym (1996) and they derived an upper bound $2.9 M_\odot$, employing WFF88 EOS, with fiducial density $\rho_f = 4.6 \times 10^{14}\,\text{g cm}^{-3}$.

Steiner et al. (2010) determined an empirical dense matter EOS from a heterogeneous data set of six neutron stars: three Type-I X-ray bursters with photospheric radius expansion and three transient low-mass X-ray binaries. Their preferred model for X-ray bursts suggests that the neutron star maximum mass is, 1.9–2.2$M_\odot$.

Corresponding to (0, 2, 0.057) and (1, 1.87, 0.057), Table 5, we found a fluid spheres having gravitational masses 1.3679$M_\odot$ and 1.3677$M_\odot$ with radius 13.21 km and moment of inertia $I_{Crab,45}$ = 1.6102 and 1.610006. These values are quite well agreement with the possible mass and radius of the Crab pulsar (> 1.2$M_\odot$, 10–14 km), using lower limit of the moment of inertia of the Crab pulsar, $I_{Crab,38}$, 1.61, for the conservative estimate of Crab Nebula mass $M_{Neb} = 2M_\odot$ (Bejger and Haensel 2002). The mass of the neutron star, in the Crab nebula, however, is most probably 1.4$M_\odot$ (MacAlpine and Satterfield 2008) with the most probable radius around 11–12 km (Steiner et al. 2010). Moreover, corresponding to the values (0, 3.76, 0.083) and (1, 3.47, 0.083) our model gives a fluid spheres of gravitational masses 1.9645$M_\odot$ and 1.9644$M_\odot$ with radius 14.38 km. These values are also quite well agreement with the predicted mass and radius of the Crab pulsar (> 1.9$M_\odot$, 14–15 km), using another lower limit, for $I_{Crab,38}$, 3.04, for the newest estimate of the Crab nebula mass, 4.6 ± 1.8$M_\odot$ (Bejger and Haensel 2002).

From Table 5, we observe that, our model yields the moments of inertia, corresponding to (0, 9.55, 0.053), (1, 9.07, 0.053) and (0, 8.43, 0.05), (1, 8.02, 0.05) for the pulsar PSR J0737-3039A and PSR J0737-3039B which lie within the range ($I_{A,38}$ = 1.4–1.7 and $I_{B,38}$ = 1.3–1.6) calculated by the EOSs which involve hyperons and/or exotic phases of dense matter (quark and kaon condensate) at supranuclear densities (Bejger et al. 2005; Worley et al. 2008).

Güver et al. (2010) reported the mass and radius of the neutron star in low-mass X-ray burster 4U 1820-30 were (1.58 ± 0.06)$M_\odot$ and (9.1 ± 0.4) km. Corresponding to (0, 12.6, 0.061) our model yields 1.5801$M_\odot$ with radius 12.99 km. But for the values (0, 11.6, 0.051) we find $M = 1.3002M_\odot$ with radius $R$ = 12.49 km which are quite similar mass, $M = (1.3 ± 0.6)M_\odot$, and radius, $R = 11^{+3}_{-2}$ km, of that neutron star (in X-ray burster 4U 1820-30) reported recently by Kuśmierek and others (Kuśmierek et al. 2011).

Table 2, 3, 4 show that our solution satisfies all the necessary physical conditions giving us a possibility for different charge variations within the fluid sphere. In absence of the charge, ($K = 0$), however, we are left behind with the static neutral Durgapal IV solution (component of exact metric, $g_{00}$, is same for $Q = 0$ and $Q > 0$), which satisfies the all physical boundary conditions (Delgaty and Lake 1998; Kiess 2012).